\documentclass{egpubl}

\usepackage{fancyhdr}
\fancypagestyle{firstpage}{
  \fancyhf{}
  \fancyhead[L]{\footnotesize \parbox[t]{0.95\textwidth}{\textit{Author’s Accepted Version. Published in \textit{Computer Graphics Forum}. Final version available at: \url{https://doi.org/10.1111/cgf.70098}}}}
  
}

\SpecialIssuePaper         

\usepackage[T1]{fontenc}
\usepackage{dfadobe}

\usepackage{cite}  
\BibtexOrBiblatex
\electronicVersion
\PrintedOrElectronic
\ifpdf \usepackage[pdftex]{graphicx} \pdfcompresslevel=9
\else \usepackage[dvips]{graphicx} \fi

\usepackage{graphicx}   
\usepackage{amsmath}    
\usepackage{mathtools}  
\usepackage{hyperref}   
\usepackage{tabularx}   

\usepackage{array}      
\usepackage{booktabs}   
\usepackage{multirow}   

\usepackage{enumitem}   
\setlist[itemize]{noitemsep, topsep=0pt} 

\usepackage{listings}   
\usepackage{xcolor}     

\definecolor{mycolor}{RGB}{33, 150, 243} 

\newcommand{\sys}{\textsc{DataWeaver}}

\usepackage{xspace} 
\newcommand{\ie}{{i.e.,}\xspace}
\newcommand{\eg}{{e.g.,}\xspace}

\newcommand{\myast}{$\ast$\xspace}

\usepackage{xcolor} 
\definecolor{lightpink}{RGB}{237,157,202}
\definecolor{lightred}{RGB}{210,121,121}
\definecolor{lightorange}{RGB}{230,170,50}
\definecolor{lightgold}{RGB}{210,194,121}
\definecolor{lightgreen}{RGB}{121,210,121}
\definecolor{lightaqua}{RGB}{121,206,210}
\definecolor{lightblue}{RGB}{121,124,210}
\definecolor{lightpurple}{RGB}{153,102,255}
\definecolor{red}{RGB}{178,34,34}
\definecolor{gray}{RGB}{166,166,166}
\definecolor{peach}{RGB}{255,218,185}
\definecolor{V2T}{HTML}{9370DB}
\definecolor{t2v}{HTML}{1B7F79}
\definecolor{T2V}{HTML}{1F4CFF}
\definecolor{component}{HTML}{5D5FEF}
\definecolor{Amber}{RGB}{255,190,11}
\definecolor{Pantone}{RGB}{251,86,7}
\definecolor{Rose}{RGB}{255,0,110}
\definecolor{Violet}{RGB}{131,56,236}
\definecolor{Azure}{RGB}{58,134,255}
\definecolor{mypurple}{RGB}{131,56,236}
\definecolor{myred}{RGB}{251,86,7}
\definecolor{mydarkred}{RGB}{252,102,32}
\definecolor{mydarkpurple}{RGB}{144,77,238}

\usepackage{soul} 
\usepackage{xargs} 

\newcommandx{\yu}[2][1=]{%
    \textcolor{orange}{\setul{}{1pt}\ul{#1} [\textbf{Yu:} #2]}%
}

\newcommandx{\vidya}[2][1=]{%
    \textcolor{purple}{\setul{}{1pt}\ul{#1} [\textbf{Vidya:} #2]}%
}

\newcommandx{\denny}[2][1=]{%
    \textcolor{teal}{\setul{}{1pt}\ul{#1} [\textbf{Denny:} #2]}%
}

\newif{\ifhidecomments}
\hidecommentsfalse
\ifhidecomments
    \newcommand{\Denny}[1]{}
    \newcommand{\Vidya}[1]{}
\else
    \newcommand{\Denny}[1]{\textbf{\small\sffamily{\textcolor{blue}{[#1 -- Denny]}}}}
    \newcommand{\Vidya}[1]{\textbf{\small\sffamily{\textcolor{brickred}{[#1 -- Vidya]}}}}
\fi

\usepackage{soul} 

\definecolor{added}{RGB}{0, 0, 0}  

\definecolor{removed}{RGB}{220, 20, 60} 
\definecolor{modified}{RGB}{255, 140, 0} 
\definecolor{comment}{RGB}{34, 139, 34} 

\newcommand{\add}[1]{\textcolor{added}{#1}}

\usepackage{egweblnk}

\title{\sys{}: Authoring Data-Driven Narratives through the Integrated Composition of Visualization and Text}

\author[Y. Fu, D. Bromley \& V. Setlur]
{\parbox{\textwidth}{\centering Y.\,Fu$^{1}$\orcid{0000-0001-5076-6299}, 
D.\,Bromley$^{2}$\orcid{0009-0007-0303-8062} and V.\,Setlur$^{2}$\orcid{0000-0003-3722-406X}} \\
{\parbox{\textwidth}{\centering $^1$Georgia Institute of Technology, Atlanta, GA, USA  \hspace{0.5cm}
$^2$Tableau Research, Seattle, WA and Palo Alto, CA, USA}}
}

\makeatletter
\let\p@copyrightTextShortEven\relax
\let\p@copyrightTextShort\relax
\let\@copyrightTextShort\relax
\makeatother

\begin{document}



\maketitle
\thispagestyle{firstpage}

\begin{abstract}
Data-driven storytelling has gained prominence in journalism and other data reporting fields. However, the process of creating these stories remains challenging, often requiring the integration of effective visualizations with compelling narratives to form a cohesive, interactive presentation. To help streamline this process, we present an integrated authoring framework and system, \sys, that supports both visualization-to-text and text-to-visualization composition. \sys{} enables users to create data narratives anchored to data facts derived from ``call-out'' interactions, i.e., user-initiated highlights of visualization elements that prompt relevant narrative content. In addition to this ``vis-to-text'' composition, \sys{} also supports a ``text-initiated'' approach, generating relevant interactive visualizations from existing narratives. Key findings from an evaluation with 13 participants highlighted the utility and usability of \sys{} and the effectiveness of its integrated authoring framework. The evaluation also revealed opportunities to enhance the framework by refining filtering mechanisms and visualization recommendations and better support authoring creativity by introducing advanced customization options.

\ccsdesc[500]{Human-centered computing~Visualization systems and tools}
\ccsdesc[500]{Human-centered computing~Interactive systems and tools}

\printccsdesc

\end{abstract}


\section{Introduction}
Data-driven storytelling, a blend of visualization, interactivity, and narrative techniques, serves as a powerful medium for effectively communicating complex data insights, engaging the public, and delivering clear and actionable information~\cite{riche2018data, Lee_more_than, Fu_2023_Morethan}. By transforming data into compelling and often interactive narrative visualizations~\cite{Segel2010NarrativeData, Hullman_2011_narrativevis}, data journalists make complex insights both understandable and impactful. 

Despite its potential, authoring data-driven stories poses an interdisciplinary challenge, requiring a combination of skills in data analysis, visualization, and narrative writing~\cite{riche2018data, Lee_more_than}. Prior work has introduced tools to streamline the process, supporting chart creation and annotation, identifying data facts, and storytelling. Yet, the diversity in authors' skillsets and workflows, coupled with the complexity of the process~\cite{Lee_more_than}, poses challenges to achieving an efficient and integrated authoring experience. First, composing data-driven narratives is often a tedious and error-prone process, requiring manual transcription of data into textual facts and, in turn, transforming these facts into coherent narratives~\cite{Chen_2022_CrossData, Fu_UIST24_data_fact_checking}. Second, current authoring tools typically follow a unidirectional `data-first' workflow, offering limited entry points for traditional authors who often begin with a narrative perspective and then seek data to refine and expand their story~\cite{Fu_2023_Morethan}. Third, the storytelling process involves context-switching between data analysis, visualization creation, and narrative writing, making it challenging to achieve a cohesive and integrated experience.

Drawing from a formative study with domain experts and existing work, our work proposes a framework~(\autoref{fig:teaser}) for an integrated visualization-text composition for data-driven storytelling, implemented in a system, \sys. The system supports \textit{vis-to-text} composition by incorporating deictic referencing through chart interactions, enabling users to create charts, highlight visual elements, and anchor Large Language Model (LLM)-based narratives to selected data facts, ensuring the narratives are aligned with the authors' intent. Complementing this approach, \sys{} also enables \textit{text-to-vis} composition by recommending and generating interactive visualizations based on selected text, helping authors jumpstart relevant data exploration. This dual approach addresses distinct but complementary needs; \textit{vis-to-text} helps ground narratives in accurate, data-driven insights, reducing manual transcription and error, while \textit{text-to-vis} supports narrative-first workflows, particularly valuable for authors with limited data expertise by providing tailored visualization recommendations driven by text. We evaluated \sys{}'s utility and usability with $13$ participants and conducted a week-long diary study with $5$ of them. Participants provided valuable feedback for refining the system, particularly regarding the data facts filtering mechanism, chart recommendations, and customization options.

\section{Related Work}
Our work draws on research in data-driven storytelling, chart and dashboard authoring, and natural language interfaces, contributing to the broader goal of effectively communicating data insights.

\subsection{Data-driven Storytelling and Narrative Visualization}
Data-driven storytelling has gained prominence as a visualization and HCI research focus, with prior work addressing key aspects such as insight discovery~\cite{Fu2022SupportingVisualization, Wang_2020DataShot, Shi_2021_Colliope}, streamlining the authoring process~\cite{Satyanarayan2014AuthoringEllipsis, Chen_2022_CrossData, Sultanum2021LeveragingVizflow, Sultanum2023DataTales:Articles, Conlen2021IdyllArticles, latif2021kori}, diversifying presentation mediums~\cite{Lee_2013_SketchStory, amini2016authoring, bach2017emerging, Cao_2023_DataParticles, Shen2023DataInterplay, Shin_2023_Roslingifier}, exploring the design space~\cite{Yang_2022_data_story_structure, Lan2022NegativeStories, LanKineticharts:Design}, and developing taxonomies and frameworks ~\cite{Li2024WhereCollaboration, Chen_2020_StorySynthesis, zhao2023stories, Chevalier2018}. 
With the recent rise in LLMs, a key focus has been on integrating AI technologies into storytelling tools, harnessing the complementary strengths of both humans and AI~\cite{Li2024WhereCollaboration}. Our work aligns with this direction, placing particular emphasis on facilitating a more efficient, fluid, and accurate authoring experience.

Several tools focus on bridging the gap between complex data analysis and accessible storytelling. For example, Ellipsis~\cite{Carr2014EurographicsEllipsis} facilitates narrative visualization through user-specified transitions and annotations, while Idyll Studio~\cite{Conlen2021IdyllArticles} provides a structured environment for authoring interactive articles that integrate text and visualizations. These tools, however, largely focus on pre-defined workflows, limiting flexibility for iterative or exploratory storytelling. Kori~\cite{latif2021kori} offers an interactive platform for synthesizing text and charts, enabling users to create coherent narratives by effectively combining these elements. CrossData~\cite{Chen_2022_CrossData} enhances this capability by leveraging connections between text and data, streamlining the authoring process of data documents.
Similarly, VizFlow~\cite{Sultanum2021LeveragingVizflow} supports dynamical layouts by leveraging text-chart links. However, bidirectional authoring support in these tools is limited to basic linkages between existing text and charts, lacking mechanisms to dynamically generate context-specific insights by leveraging their inherent connections.

The introduction of block-based approaches, such as DataParticles~\cite{Cao_2023_DataParticles}, adds another dimension to storytelling by enabling dynamic and animated data representation. The exploration of LLMs in tools, such as DataTales~\cite{Sultanum2023DataTales:Articles} further pushes the boundaries of narrative generation, though it presents challenges with scalability and accuracy with respect to the generated summaries. Charagraph~\cite{Masson2023Charagraph:Paragraphs} facilitates real-time annotation and narrative generation by allowing users to create interactive charts within data-rich paragraphs, enhancing the narrative process. Meanwhile, ChartAccent~\cite{RenChartAccent:Storytelling} provides tools for annotating data visualizations, enabling users to highlight key insights and make data stories more comprehensible. Automation also plays a crucial role in advancing data-driven storytelling. Calliope~\cite{Shi_2021_Colliope} automates the generation of visual data stories from spreadsheets, translating raw data into narrative presentations. Notable~\cite{Li2023Notable:Notebooks} streamlines the storytelling process within computational notebooks, allowing for the seamless integration of narratives with analytical workflows. Lastly, Erato~\cite{Sun2022Erato:Interpolation} introduces a collaborative aspect by enabling cooperative editing of data stories through fact interpolation, facilitating a shared narrative construction process among multiple users. Although these tools advance the capabilities of data-driven storytelling, they often focus on the linear conversion of data into stories. Our work further enhances the storytelling process by employing a novel framework that integrates both vis-to-text and text-to-vis workflows. 

\subsection{Chart and Dashboard Authoring}
In data visualization and dashboard authoring, several tools aim to enhance the expressiveness of visual presentations. Medley~\cite{Pandey2022MEDLEY:Composition} provides intent-based recommendations to support the composition of dashboards. Data Illustrator \cite{Liu2018DataAuthoring} and Data Animator \cite{Thompson2021DataGraphics} augment traditional tools by integrating data binding functionalities, enabling users to directly link datasets to graphical elements for dynamic visualization design. Beyond traditional visualization tools, innovations in data content authoring further expand the scope of data storytelling. Epigraphics~\cite{Zhou2024Epigraphics:Authoring} introduces a message-driven approach to infographics authoring, focusing on the narrative elements that drive the creation of infographics and conveying data insights in a more expressive way.
DataInk~\cite{Xia2018Dataink:Drawing} provides a creative platform for direct data-oriented drawing, allowing users to create visualizations through an intuitive drawing interface. By focusing on creativity and direct manipulation of data, DataInk enables users to explore and present data in innovative and personalized ways. These authoring tools often focus on either visualization or narrative as the starting point; however, \sys{} enables dynamic transitions between visualization creation and narrative development, anchoring narratives to data facts and visualization recommendations.

\subsection{Natural Language Interfaces}
The development of natural language interfaces (NLIs)~\cite{Setlur2016Eviza:Analysis,analyza, datatone,orko} has enabled users to engage with data through conversational interactions, simplifying complex visual analysis by enabling natural language queries to explore and interpret data. 
NL4DV~\cite{Narechania_2021_NL4DV} maps natural language queries to JSON-based analytic specifications with data attributes, tasks, and Vega-Lite visualizations. 
Voder~\cite{Srinivasan_2019_Voder} enhances data interpretation by making textual data facts interactive for exploring alternative visualizations.
FlowSense~\cite{Yu2020FlowSense:System} integrates natural language processing with visual data exploration, allowing users to construct and modify visualizations via conversational input within a dataflow system. BOLT~\cite{Srinivasan2023BOLT:Authoring} enhances dashboard authoring by allowing users to create and modify dashboards using conversational interactions, minimizing the need for manual configurations. 
While existing tools for data-driven storytelling, chart authoring, and natural language interfaces offer various capabilities, they often lack a fully integrated approach that supports fluid transitions between data and narrative. Many focus on either visualization or narrative as the primary entry point, limiting the flexibility of authors in switching between different modes of creation. \sys{} addresses these gaps by providing a bidirectional storytelling environment that supports both \textit{data-first} and \textit{text-first} authoring modes. The system provides a novel, flow-based, zoomable interface that enables navigation between visualizations and narratives, supporting the integration of data and text during the storytelling process.

\section{Understanding the Challenges in Data-driven Storytelling}
Prior research highlights the complexity of data storytelling, which often involves iterative data exploration, the composition of text and visuals, story arrangement, and final presentation~\cite{Lee_more_than}. Our work aims to help streamline this process by integrating visualization and text authoring into a cohesive workflow. To uncover key obstacles in achieving this integrated experience, we conducted a formative study involving a data story authoring exercise and post-study interviews with five professional visualization practitioners ($P'_1$–$P'_5$) recruited via our organization's mailing list. All participants self-identified as experts with extensive experience in creating visualizations and data-driven stories.
%
In the authoring exercise, we asked participants to explore data and uncover insights using multiple interactive visualizations, preferably their own creations; for those who did not have their own creations, we provided a curated list of visualizations sourced from Tableau Public~\cite{tableaupublic} and The New York Times~\cite{NYT} (included in supplementary material). They presented their findings by capturing screenshots, drafting textual narratives, and adding annotations to create a cohesive presentation. Throughout the process, we encouraged participants to think aloud and share their reasoning. Each session lasted 30–40 minutes, followed by a 10–15 minute semi-structured interview. The study allowed us to observe authoring processes, generate artifacts for analysis, and gather feedback on interaction methods, challenges, and perspectives on AI integration. Sessions were recorded and transcribed, and the generated stories were collected using Google Slides.
From our observations, transcripts, and collected stories, we identified the following key findings: 

\noindent\textbf{\textit{F1. Pre-computed data facts can be helpful.}} Three participants expressed a desire for pre-calculated data facts describing statistics computed from relevant data points as they interacted with visualizations. $P'4$ noted, \textit{``Having some words already available to be plugged in without writing the calculation, like the percentage change, would be fantastic.''} $P'3$ concurred, expressing a desire for the calculation of the \textit{``absolute value''} of an increasing trend.

\noindent\textbf{\textit{F2. Challenges in manual data transcription from visualizations.}} Four participants highlighted the challenges of manually transcribing numerical data from visualizations into text. $P'4$ described this as \textit{``[The hardest part] was getting some of the numbers from the visual into the text...so I had to go back and forth.''} This issue is exacerbated when values are not directly visible, requiring interactions like hovering to reveal them. For instance, in a chart showing score differences, $P'1$ needed the exact value (\eg \textit{25-point lead}) to write an accurate narrative. 

\noindent\textbf{\textit{F3. Concerns regarding direct input of raw data to AI.}} While three participants showed a strong willingness to use AI, particularly LLMs for data narrative composition, four participants voiced significant concerns about feeding raw datasets directly to AI. They cited concerns over \textit{accuracy} of generated narratives and \textit{data security}, emphasizing their reluctance to expose sensitive data to AI.
   
\noindent\textit{\textbf{F4. Narrative-guided exploration as an authoring strategy.}} We observed that authors working with familiar datasets or topics often adopt a narrative-driven approach---rather than engaging in open-ended data analysis, they begin with a specific storyline in mind, actively seeking data points or visual facts to substantiate or refine their narrative. For instance, $P'_1$, while exploring sports data, already had a clear narrative direction based on their personal experience of watching a specific game; they intentionally identified data facts to enrich and help validate their pre-existing narrative.
 
\noindent \textbf{Design Goals:} Drawing from these key findings and prior research~\cite{Lee_more_than, Chen_2022_CrossData, Fu_2023_Morethan}, we distill five design goals for devising an integrated authoring framework to help streamline both visualization and text composition in data-driven narratives:
{\small
\begin{itemize}
    \item \textbf{DG1}. Provide accurate data insights/facts that align with user intentions.
    \item \textbf{DG2}. Streamline data transcription and narrative composition.
    \item \textbf{DG3}. Prevent the direct input of raw datasets into LLMs.
    \item \textbf{DG4}. Enable narrative-driven visual data exploration.
    \item \textbf{DG5}. Foster a seamless and cohesive authoring experience.
\end{itemize}
}

\section{\sys{} System}

\begin{figure*}[t]
 \centering
 \includegraphics[width=0.95\linewidth]{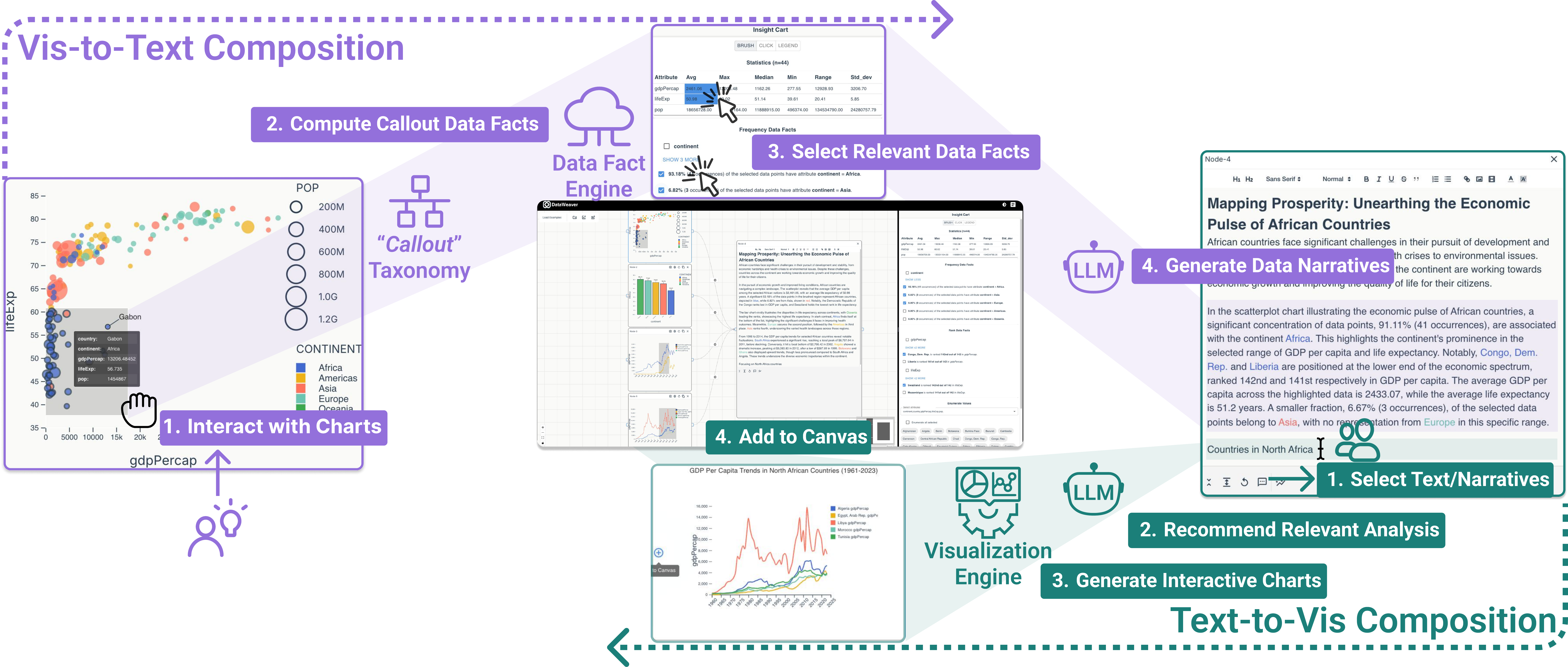}
 \caption{
 The integrated framework underlying \sys. The interface is detailed in \autoref{fig:system_interface}. 
 The \textcolor{V2T}{\textbf{purple-colored flow}} shows the \textit{Vis-to-Text} composition (\autoref{fig:vis-to-text}).
 The \textcolor{t2v!80}{\textbf{teal-colored flow}} demonstrates \textit{Text-to-Vis} composition (\autoref{fig:Text-to-vis})
 }
 \label{fig:teaser}
\end{figure*}

\subsection{An Integrated Framework for Visual-Text Composition}
To achieve our design goals, we propose an integrated framework (\autoref{fig:teaser}) for data story authoring, which combines two compositional approaches: \textit{vis-to-text} and \textit{text-to-vis}.

\textbf{\textit{Vis-to-Text Composition}} addresses design goals \textbf{DG1} and \textbf{DG2} by facilitating narrative generation anchored to authors' \textit{intended} insights and \textit{accurate} data facts. Specifically, we propose the following approach. First, we allow users to \textit{specify visual elements of interest through chart interactions}. These interactions enable deictic referencing by highlighting elements to indicate that the elements form the focus of the discussion. Second, we rely on algorithms to \textit{compute descriptive statistics and data facts for the highlighted data points} through an intermediate data fact layer, ensuring accurate and efficient computation by delegating tasks to algorithms instead of LLMs. The feature also enables users to select relevant data facts to enhance narrative relevance and anchors generated narratives to a defined set of data facts to improve accuracy. Finally, we \textit{leverage LLMs to generate narratives anchored to the selected data facts}. By weaving pre-computed data facts into narrative text, we address challenges such as restricted context length, limited numerical precision, and difficulties with complex data operations, avoiding the need to feed raw data directly into LLMs, fulfilling \textbf{DG3}, and offering an additional layer of privacy control.

\textbf{\textit{Text-to-Vis Composition}} addresses \textbf{DG4} by introducing text as an alternative starting point for data story creation. A common challenge in storytelling is the need to create visualizations to support further analysis or to expand existing narratives. By simplifying this process, we aim to streamline authoring and reduce the burden on users with limited data expertise. To achieve this, the generation of relevant visualizations from text requires a degree of inter-modal semantic mapping. However, the free-text nature of \sys{} was not amenable to a strictly structured parsing approach. Given these design considerations, we leverage LLMs as a pragmatic solution, offering a reasonable balance between semantic interpretation and free-text comprehension. While not without limitations, we found that LLMs provide an effective means of suggesting relevant analyses and generating chart specifications under the constraints of unstructured textual input. 
This approach also complements \textit{vis-to-text} composition, as the resulting interactive visualizations can serve as new starting points for further data exploration and narrative development. Together, these bidirectional workflows create a `boomerang-like' loop, enabling a seamless and iterative authoring experience that addresses \textbf{DG5}.

\subsection{Callout Taxonomy}

\begin{table}[b]
\centering
\scriptsize
\renewcommand{\arraystretch}{1.0} 
\setlength{\tabcolsep}{2pt} 
\begin{tabularx}{\columnwidth}{m{1cm} m{2cm} m{5cm}}
\toprule
\textbf{Chart} & \textbf{Interaction} & \textbf{Callout Data Facts} \\
\midrule
\multirow{4}{*}{\centering Scatterplot}
 & Area Selection \newline (2-D Brush) & Summary statistics\myast, Frequency\myast, Group vs. Global (stats), Rank\myast, Values\myast \\
 & Discrete Selection \newline (Click) & Values\myast, Outliers\myast, Rank\myast, Summary statistics\myast, Group vs. Global, Frequency\myast \\
 & Group Selection \newline (Legend Click) & Summary statistics\myast, Frequency\myast, Rank\myast, Group vs. Global (one group selected), Group vs. Group (multiple groups), Outliers\myast, Values\myast \\
 & Add Trendline & Correlation between variables, Trendline \\
\midrule
\multirow{3}{*}{Bar Chart} 
 & Discrete Selection \newline (Click) & Values\myast, Rank/Extreme\myast, Difference\myast, Summary statistics\myast \\
 & Category Selection \newline  (Legend Click) & Summary statistics\myast, Frequency\myast, Rank\myast, Group vs. Global Difference, Group vs. Group Difference, Outliers\myast, Values\myast \\
 & Area Selection \newline (1-D Brush) & Summary statistics\myast, Frequency\myast, Rank\myast, Group vs. Global (one group), Values\myast \\
\midrule
\multirow{3}{*}{Line Chart} 
 & Timeframe Selection \newline (Brush) & Trend\myast, Start/end time\myast, Extreme points\myast, Range\myast, Difference\myast \\
 & Line Selection \newline (Legend Click) & Trend\myast, Compare selected lines\myast, Compare selected to others \\
 & Temporal Point \newline Selection (Click) & Value/Rank for individuals, Compare selected date, Analyze position relative to trend \\
\midrule
\multirow{2}{1cm}{Stacked Bar Chart} 
 & Subcategories Selection (Legend Click) & Comparison/relation/joint contribution, Joint relationship to others, Comparison to each other, Compare individuals to group \\
 & Segments Selection \newline  & Proportions, Compare individual proportions \\
\midrule
Donut/Pie Chart 
 & Discrete Selection \newline  (Click) & Proportional values, Compare selected proportions, Compare sum proportion to total \\
\midrule
\multirow{2}{*}{Sunburst} 
 & Discrete Selection \newline (Click) & Proportional values, Compare relative proportions, Compare/aggregate hierarchical proportions \\
 & Chained Selection \newline (Double-click) & Chained proportion \\
\bottomrule
\end{tabularx}
\caption{\textit{Callout} Taxonomy categorizes chart types, their corresponding callout interaction types, and the associated callout data facts. A \myast denotes a calculation implemented in \sys.}
\label{callout_intent}
\end{table}

\begin{figure*}[htb]
  \centering
  \includegraphics[width=0.93\textwidth]{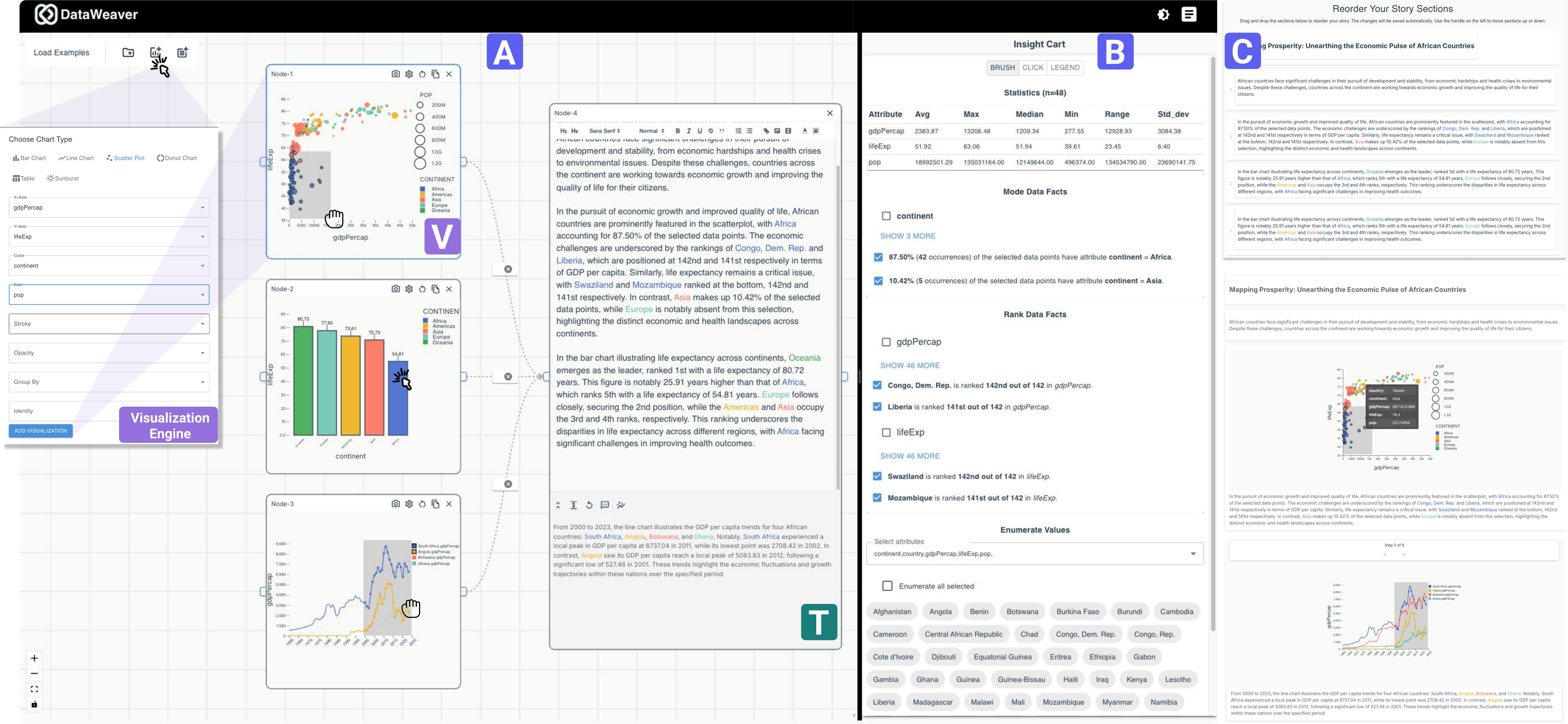}
  \caption{An overview of \sys's interface. The \textit{\textbf{Authoring Canvas}}~\colorbox{component!90}{\textbf{\textcolor{white}{A}}} is a flow-based, zoomable interface where users can add \textit{vis-nodes} (\colorbox{V2T!90}{\textbf{\textcolor{white}{V}}}) using a visualization engine and \textit{text-nodes} (\colorbox{t2v!80}{\textbf{\textcolor{white}{T}}}), as well as edges to connect them. The \textit{\textbf{Insight Cart}} \colorbox{component!90}{\textbf{\textcolor{white}{B}}} is used for insight management while the \textbf{\textit{Review Page}}~\colorbox{component!90}{\textbf{\textcolor{white}{C}}} allows users to reorder the story and convert the story pieces into different presentation formats.   
  }
  \label{fig:system_interface}
\end{figure*}

Given the extensive range of statistics and data facts that can be derived from a dataset~\cite{Shi_2021_Colliope}, 
a key challenge for our proposed framework lies in striking a balance between providing users with granular control to define \textit{`what is salient'}, while minimizing their cognitive effort in identifying and selecting relevant data facts.

Achieving this balance requires an effective mechanism for computing and organizing data facts---one that clarifies user intent and generates a focused list of insights based on interactions. User intent behind interactions has been studied across various contexts~\cite{Yi2007TowardVisualization, Snyder2024DIVI:Visualization}, such as analysts' selection behaviors in scatterplots~\cite{GadhavePREDICTINGVISUALIZATIONS} and annotations in grouped bar charts~\cite{rahman2024exploring}. We extend this understanding by examining intent behind \textit{``calling out''} visual elements, i.e., \textit{``When users highlight data points in a chart, what insights or facts are they likely to incorporate into their narrative or discussion?''} We developed a preliminary \textbf{\textit{Callout} Taxonomy} (\autoref{callout_intent}) as a theoretical guidance for more relevant data facts generation and their effective organization. Drawing from prior literature on visualization interaction~\cite{Amar_2005_low_level_analytic_activity, Yi2007TowardVisualization}, data facts and insights~\cite{Wang_2020DataShot, Shi_2021_Colliope}, and analytical intent~\cite{GadhavePREDICTINGVISUALIZATIONS, Snyder2024DIVI:Visualization}, we posit that the intended \textit{data facts} are associated with the \textit{chart type} and \textit{interaction type} that users engage with. In this context, we define \textit{Callout Interactions} as user actions that highlight specific data points in a visualization, \textit{Callout Intent} as the motivation behind such highlights, and \textit{Callout Data Facts} as the insights derived from these interactions.
We selected a set of exemplary chart types, both standard and advanced, and curated a list of \textit{callout interactions} for each, drawing from popular visualization tools/libraries, and prior research~\cite{Heer_2008_generalized_selection}. Two authors collaboratively developed the corresponding \textit{callout data facts}: one created the initial coding, while the other independently generated a separate set. The lists were iteratively refined into a final taxonomy, extendable to additional chart types, interactions, or even compounded and chained interactions.

\subsection{User Interface}

Building on this taxonomy, we integrated its principles into \sys{} to inform the design and functionality of the system's interface (\autoref{fig:system_interface}) and core features. The frontend is built with React and uses D3.js for rendering interactive visualizations. \sys{}'s backend, built with Python Flask, handles server-side logic and API requests, utilizing GPT-4o as its LLM along with other libraries for text generation, semantic parsing, and various computational tasks.
The interface comprises three major components: a flow-based \textit{Authoring Canvas} (\autoref{fig:system_interface}-A) for composing visualizations and text, an \textit{Insight Cart}~(\autoref{fig:system_interface}-B) for managing insights, and a \textit{Review Page} (\autoref{fig:system_interface}-C) for refining the story and previewing the final visual data narrative in multiple formats.

\textbf{Authoring Canvas (\autoref{fig:system_interface}-A):}
The authoring canvas allows users to add two types of nodes: visualization nodes (\textit{vis-nodes}) and text editor nodes (\textit{text-nodes}). Edges connect these nodes, defining their relationships and flows.

\setlength{\leftmargini}{0pt}
\setlength{\itemindent}{0pt}
\begin{itemize}[leftmargin=*, labelindent=2pt, style=unboxed]
    \item \textbf{\textit{Nodes:}}  
    Users can add various types of interactive visualizations as \textit{vis-nodes}. \sys{} fundamentally supports interactive visualization creation by leveraging a visualization engine powered by D3.js. 
    Users can upload and manipulate datasets (\eg filtering), select the chart type, and specify the attributes that determine the chart's visual encodings. The chart types supported are congruent with \autoref{callout_intent}. The embedded visualizations are interactive, featuring tooltips and corresponding callout interactions. \textit{Text-nodes} provide a straightforward, rich text editor for narrative composition, along with widgets for additional text manipulation. Users can also move, zoom, resize, duplicate, and remove nodes from the canvas. 
    \item \textbf{\textit{Edges:}}  
    Edges help specify the flow of data facts generated by chart interactions. For example, if \textit{vis-node-1} and \textit{vis-node-2} both have edges directed to \textit{text-node-3}, the data facts provided by these two visualization nodes will collectively contribute to the text-node. Users can reuse the data facts from the same nodes to drive different text-nodes. This mechanism aims to assist users in managing story content and flow, especially when composing a data story that involves multiple sets of data facts and visualizations. Datasets bound to vis-nodes will also automatically serve as source data for visualization recommendations. Similarly, users can serve the datasets as data tables without specifying the chart. 
\end{itemize}

\begin{figure*}[t]
  \centering
  \includegraphics[width=0.92\textwidth]{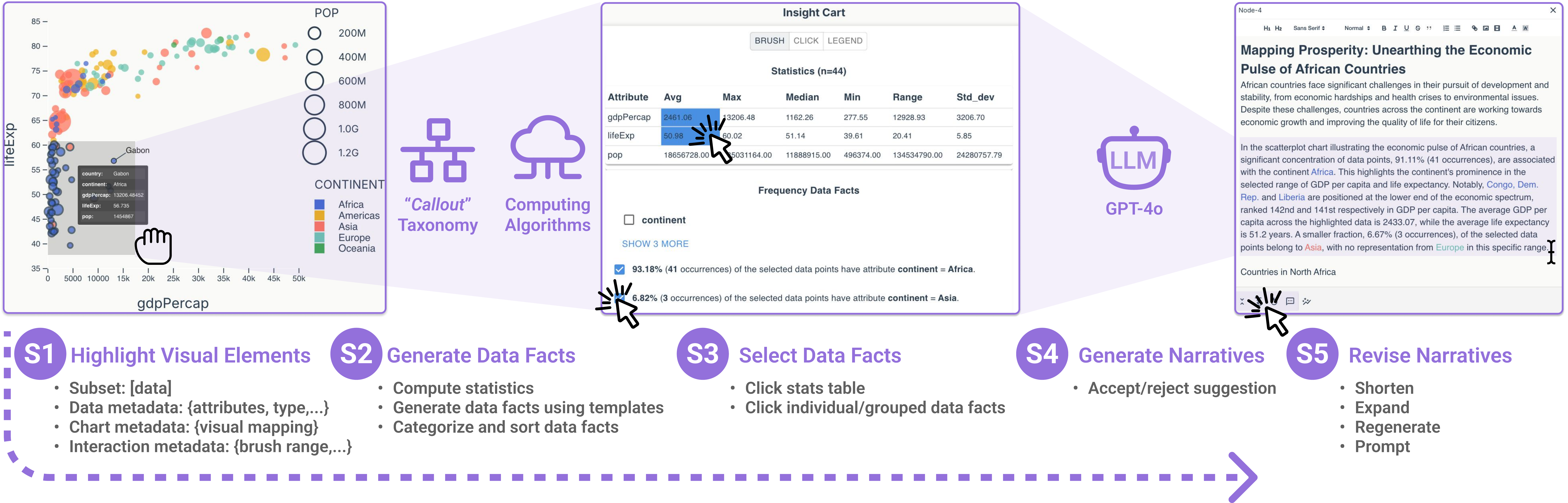}
  \caption{Demonstration of Vis-to-Text composition workflow. After users apply callout interaction to visualization~(S1), \sys{} computes the data facts and presents~(S2) them in the insight cart. Users then select desired data facts~(S3). An LLM then generates data narratives~(S4) based on the selected data facts and metadata. Users can revise the generated narratives using the buttons~(S5).}
  \label{fig:vis-to-text}
\end{figure*}

\textbf{Insight Cart (\autoref{fig:system_interface}-B):}
The insight cart, like a ``shopping cart,'' acts as a temporary repository to store facts for ``checkout,'' \ie integration into a data story. Each node has its own insight cart.

\begin{itemize}[leftmargin=*, labelindent=2pt, style=unboxed]
    \item \textbf{\textit{Vis-nodes' Insight Carts}}: Insight carts associated with vis-nodes contain insights (e.g., descriptive statistics or data facts) derived from users' callout interactions. These insights are displayed as clickable tables and grouped checkboxes, allowing users to make selections. The selected data facts automatically flow into the insight carts of downstream \textit{text-nodes}.
    \item \textbf{\textit{Text-nodes' Insight Carts:}} Insight carts associated with text-nodes consist of two segments: a data-facts segment, which displays grouped data facts from upstream \textit{vis-nodes}, and a visual insights segment, which suggests recommended visualizations.
\end{itemize}

\textbf{Review Page (\autoref{fig:system_interface}-C):}
During or after composing the data story, authors can utilize the \textit{review page} to adjust the order of the content and preview the final output in various forms. The \textit{adjusting mode} automatically processes text from different nodes into a nested structure, enabling efficient restructuring of both the narrative and the linked visualizations. The \textit{previewing mode} allows for a preview of the final product in common narrative visualization formats: \textit{continuous page}, \textit{scrollytelling}, and \textit{stepper}.

\subsubsection{\textit{Vis-to-Text} Composition Workflow}\label{sec_core_features}

The \textit{vis-to-text} composition starts with a created visualization and follows an \textit{interact-compute-select-generate-revise} workflow with users, algorithms, and LLM collaboratively handle respective steps:

\noindent\textbf{S1. User interacts with charts to highlight visual elements:}   
In each visualization node, users initiate \textit{callout interactions} to select visual elements, prompting the system to retrieve and dispatch a package with the data subset, metadata, and interaction details. This metadata is divided into three categories: \textit{data metadata}, containing attribute names and value types; \textit{chart metadata}, which captures visual encodings like color, size, and tooltip content (e.g., mapped variables or identity variables like a country's name); and \textit{interaction metadata}, which varies depending on the chart and interaction type. For instance, brushing a scatterplot captures a 2D spatial selection with coordinate ranges $[\text{x}_1, \text{y}_1]$ to $[\text{x}_2, \text{y}_2]$ and value ranges $[\text{xValue}_1, \text{yValue}_1]$ to $[\text{xValue}_2, \text{yValue}_2]$.

\begin{figure*}[htb]
  \centering
  \includegraphics[width=0.92\textwidth]{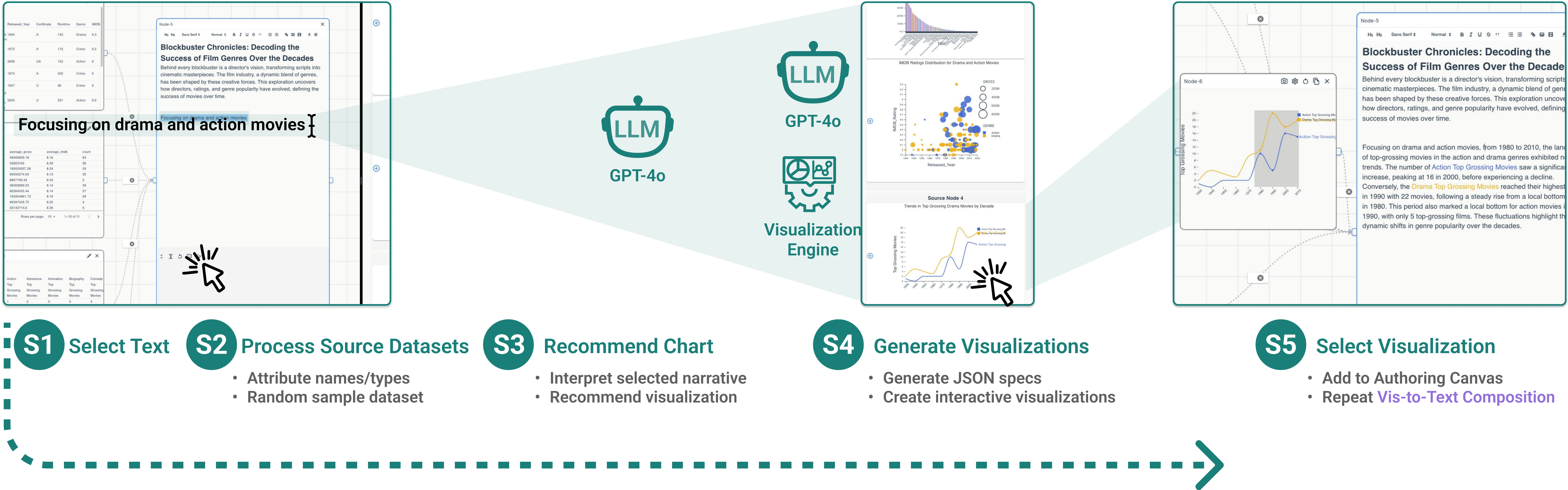}
  \caption{Demonstration of \textit{Text-to-Vis} composition workflow. Users first type or select text to focus~(S1), and \sys{} retrieves and processes the datasets from upstream nodes~(S2). An LLM then interprets the text and metadata and recommends relevant charts~(S3). Based on the charts' types, LLM then generates JSON specifications that contain both data operation and visualization schemas used to create interactive charts~(S4). Users finally review the generated charts and add desired ones as new vis-nodes~(S5).}
  \label{fig:Text-to-vis}
\end{figure*}

\noindent\textbf{S2. Algorithms compute and organize data facts:} 
Upon receiving the `callout package,' \sys's backend, following \autoref{callout_intent}, first computes the relevant statistics following the taxonomy and converts the computed values into a statistical table or template-based data facts, such as \textit{``95.24\% of the selected data points have the attribute continent = Africa''}, or \textit{``Gabon is an outlier in lifeExp.''} These data facts are presented in the \textit{vis-nodes' insight cart}. To manage this potentially extensive list of data facts and reduce users' cognitive load, \sys{} categorizes them based on derived callout intents and applies sorting mechanisms to the list. 
The data facts are organized hierarchically in a nested structure: 
\textit{Fact Types} $\triangleright$ \textit{Attributes} $\triangleright$ \textit{Data Facts}. 
The \textit{Fact Type} layer ordering is guided by both our taxonomy and the goal of maintaining UI consistency. For example, the statistical table is always placed at the top to ensure uniformity across the interface, and the \textit{Frequency} category is positioned above the \textit{Rank} category when brushing a scatterplot.
The \textit{Attribute} layer does not adhere to a specific ordering scheme. However, to reduce the computational load from handling numerous attributes and ease users' cognitive load when reviewing data facts, \sys{} processes only the \textit{attributes of interest}, \ie variables that are visually mapped and explicitly selected by the user.
Within each list, \textit{Data Facts} are sorted based on their `significance' or `interestingness,' a method frequently used by recent studies~\cite{Srinivasan_2019_Voder, Shi_2021_Colliope}. \sys{} utilizes bespoke sorting algorithms tailored to specific fact types, aiming to promote the most `significant' data facts within each category. For example, when brushing over the lower quadrant of a scatterplot visualizing life expectancy against GDP per capita, \sys{} surfaces data facts, such as \textit{``93.18\% of them are African countries''} but \textit{``none of them are European countries''} in the \textit{Frequency $\times$ Continent} category. \sys{} then sorts these \textit{frequency} data facts by calculating weighted scores, taking into account normalized differences, ratios, and entropy, while prioritizing significant deviations and prominence within the subset. \add{The detailed sorting mechanism and explanations are in the supplementary materials.}

\noindent\textbf{S3. User selects data facts:} Users click on the statistical table cells or checkboxes to select individual or grouped data facts they want to include from each visualization node. The selected data facts are then streamed to the subsequent text-nodes, where they appear in the corresponding insight cart in a nested format.

\noindent\textbf{S4. LLM generates data narratives:} 
The narrative generation can be triggered by the user pressing the \texttt{Tab} key. The selected data facts are then fed to the LLM, along with the preceding context and metadata, organized into a prompt template. The prompt template encapsulates both a generic template and chart-specific templates. 
The generic template offers a structured approach to understanding the visualization and its context. The specification includes steps such as recognizing the visualization type, examining the article context, and synthesizing data facts into a coherent narrative.
The type-specific templates provide tailored guidance and context. For example, brushing interactions provide the range of the brushed area versus the full axes range. A pseudo prompt is as follows:
{\small
\begin{itemize}[leftmargin=*, itemsep=0pt, parsep=0pt]
    \item Understand the visualization: \textit{chart type}, \textit{chart metadata}
    \item Consider the context: \textit{``preceding text content''}
    \item Consider the callout interaction: \textit{interaction metadata}
    \item Focus on these data facts: [\textit{``data fact 1''}...].
    \item Synthesize information and write a narrative based on the data facts.
\end{itemize}
}
Users can decide to accept or reject the generated narrative. Prompt templates are available in the supplementary materials.

\noindent\textbf{S5. User and LLM revise the generated data narrative.} After accepting the initial narrative, users can either manually revise it or use the LLM for further refinement. They can select any part of the generated text and prompt the LLM to revise it. \sys{} offers three shortcut buttons (\textit{shorten}, \textit{expand}, \textit{regenerate}) and a text instruction window for direct prompting. The pre-defined or user-specified instructions will be integrated into a new prompt alongside the generated content and the original prompt to ensure that the new content accurately aligns with the data facts.

\subsubsection{\textit{Text-to-Vis} Composition Workflow}
\sys{} utilizes LLMs to understand and infer contextually relevant information, enabling authors to expand their narratives using relevant information from the data. Additionally, \sys{} leverages LLMs' semantic parsing capabilities to provide a well-structured JSON specification that can be used for data operations and visualization generation~\cite{Fu_UIST24_data_fact_checking}.

\noindent\textbf{S1. User selects or types text to focus.} In a text node, users first select any portion of the existing text of interest, or they can type new sentences. For example, an author might focus on a specific region of Africa and type \textit{``Countries in North Africa...''} They then select this segment and click the \textit{Recommend Visualization} button.

\noindent\textbf{S2. Algorithm processes the datasets.} \sys{} simultaneously retrieves all the underlying datasets from upstream nodes as reference data and dispatches them for further processing, e.g., extracting attributes, aiming to provide an overview of the available datasets without uploading the entire dataset.

\noindent\textbf{S3. LLM interprets the narrative and recommends relevant charts for analysis.}  
The selected text from S1 and attribute names from S2 are fed into the LLM. The prompt guides the LLM to first interpret the available datasets using processed information from S2 and then to recommend analyses and visualizations for the selected text. For example, providing the narrative, \textit{``Women's participation in the Olympics has increased over time''} alongside a dataset of athlete counts by gender across Olympic history prompts the LLM to suggest analyses, such as the percentage of genders over the years, and a line chart to support the analysis.

\noindent\textbf{S4. LLM generates JSON specifications for data operation and visualization generation.} Based on different chart types, \sys{} prompts the LLM to generate a JSON specification for the data operations (e.g., filtering, aggregation) and visualization creation. Under the hood, \sys{} uses a JSON schema tailored for D3 to render different visualizations. For example, a scatterplot requires numerical \texttt{xAttr} and \texttt{yAttr} as mandatory inputs, while other attributes (e.g., \texttt{colorAttr}) are optional. The LLM leverages these schemas to generate corresponding JSON specifications, allowing \sys{} to create visualizations with integrated callout interactions, which are then displayed in the visual insight section of the text-node's insight cart.

\noindent\textbf{S5. User selects generated visualizations.} Eventually, users review the recommended charts and, by clicking an `add' button, select those they wish to incorporate into the canvas as new visualization nodes. These new nodes are now available for interaction, data-fact generation, and narrative inclusion in the same manner as the existing visualization nodes. The new visualization nodes have completed the bidirectional loop, allowing users to generate new data facts and narratives that are relevant to the focused text.

To summarize, \sys{} allows users to initiate the authoring process using either \textit{Vis-to-Text} or \textit{Text-to-Vis} workflows. In the former (\autoref{fig:vis-to-text}), users begin by creating interactive visualizations using \sys’s foundational chart authoring feature, then compose data narratives by interacting with the charts and the automatically generated data facts (\autoref{fig:system_interface}-B). In the latter (\autoref{fig:Text-to-vis}), users select the textual narratives they want to focus on, prompting \sys{} to recommend and generate relevant interactive charts, thus completing the loop for data-driven narrative composition. The narrative can be further refined manually or through the system's LLM-based revision feature. Finally, the visual-to-text connections automatically established enable users to rapidly adjust the sequence of content and review it in dynamic presentation formats (\autoref{fig:system_interface}-C). A video showcasing \sys's interactivity and features is included in the supplementary materials.

\section{User Evaluation}
To evaluate the effectiveness and usability of \sys{} and gather valuable feedback, we conducted a two-part evaluation: a focused user study, followed by a longitudinal week-long diary study.

\subsection{Participants and Study Protocol}
\noindent\textbf{Participant Recruitment:}  
\add{We recruited 13 participants ($P1$--$P13$) to evaluate \sys{}. Each received a \$20 Amazon gift card for a one-hour virtual session conducted via Google Meet. Participants (6 female, 7 male) self-reported high familiarity with data-driven stories (8 ``highly familiar'' and 8 ``somewhat familiar''), diverse educational backgrounds (6 PhDs, 2 master's, 4 bachelor's, 1 other), and job roles (6 data analysts, 4 students, 1 postdoc, 1 data coach, 1 software manager). Five participants took part in a subsequent diary study, each receiving a \$60 Amazon gift card.}

\noindent \textbf{User Study:} The study comprised four main phases. Participants began with an \textit{Introduction} that included a verbal overview of the study's background and objectives and a video walkthrough to ensure familiarity with \sys{} before proceeding to the tasks. This was followed by the \textit{Reproduction Walkthrough}, where participants explored \sys{}'s core functionalities by recreating a visual data story similar to the one shown in the introductory video. To streamline this process, we provided three preconfigured visualization nodes and a text node with leading narratives, and participants were encouraged to ask questions as needed. In the \textit{Open Authoring Exercise}, participants created their own visual data stories using an IMDb movie dataset~\cite{imdb}, with the option to start from four preconfigured charts or build from scratch using text linked to the datasets. We encouraged them to focus on core tasks, such as data fact selection, narrative generation, and chart recommendations, enabling us to evaluate \sys{}'s integrated framework and both composition workflows. Throughout this task, participants were encouraged to think aloud, especially when making decisions about visual elements and data facts. Finally, in the \textit{Questionnaire and Post-study Interview}, participants provided feedback on \sys{} through a Likert-scale questionnaire that included nine utility questions and ten System Usability Scale (SUS) questions~\cite{brooke1996sus}. Afterward, a follow-up interview was conducted to gather deeper insights and qualitative feedback on specific aspects of \sys{}. Detailed materials, including the questionnaire, procedure, scripts, and results, are available in the supplementary materials.

\noindent\textbf{Diary Study:}  
In the diary study, participants used \sys{} independently over a week, spending 20–30 minutes each day creating visual data stories. They took screenshots of their work, documented their experiences, challenges, and observations in daily diary entries, and provided feedback on \sys{}'s features, usability, and workflows. At the end of the week, participants shared overarching reflections and recommendations for enhancing \sys{}'s effectiveness and overall experience.

\subsection{Results and Discussion}
We analyzed responses using quantitative SUS scores and thematic coding of study transcripts. Two authors independently coded the transcripts using an open-coding approach to identify recurring themes and insights on the authoring experience, core functionalities, and areas for improvement. \add{A third author examined the diary study documents, identified new themes, and synthesized findings from both studies to form the qualitative results.}

\begin{figure}[htb]
  \centering
  \includegraphics[width=1\columnwidth]{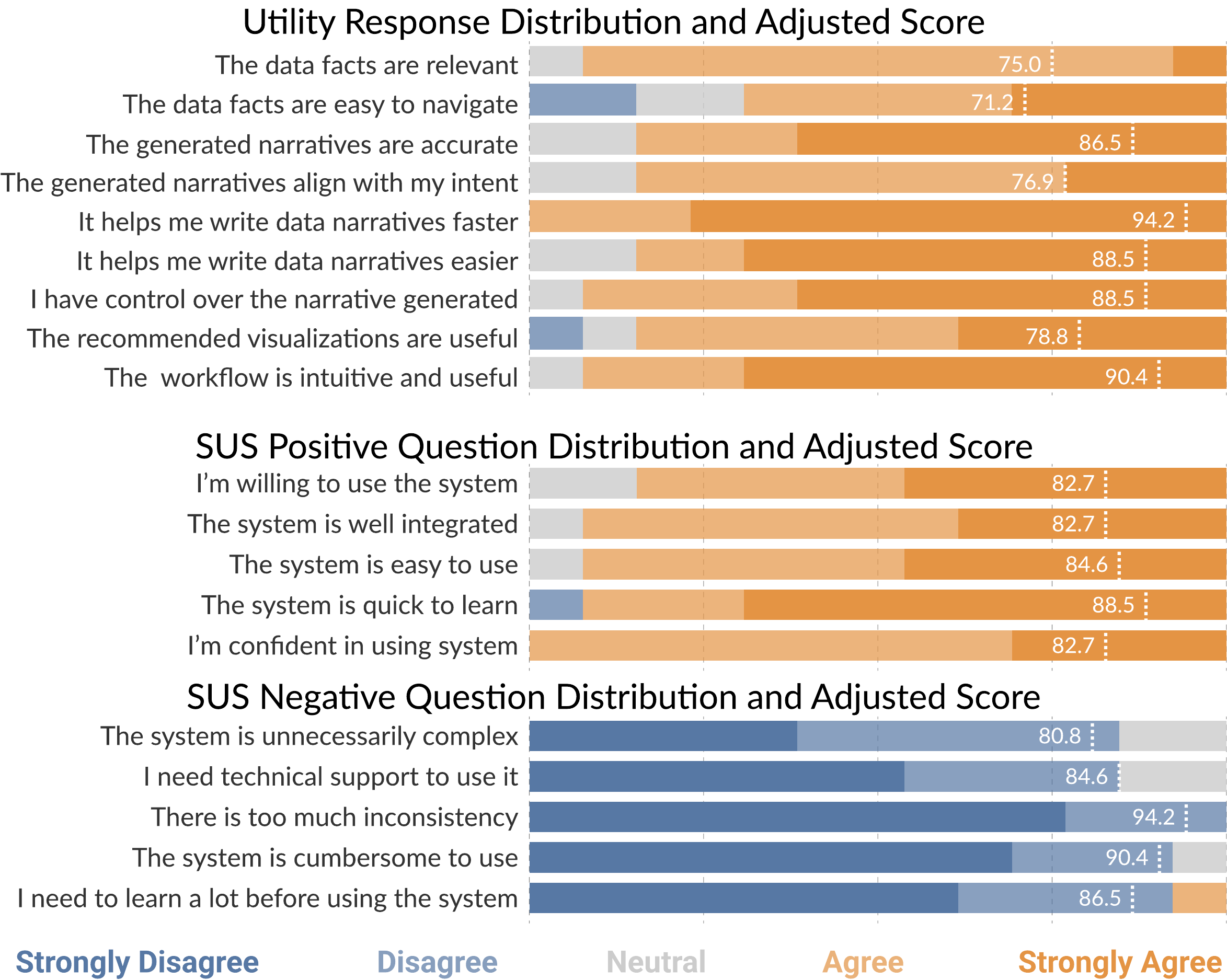}
    \caption{The charts depict questionnaire response distributions on a 5-point Likert scale. The top chart presents results of nine utility questions, while the bottom two display usability questions. White lines and numerical values represent adjusted scores (out of 100), calculated using the SUS algorithm~\cite{brooke1996sus}.}
  \label{fig:distribution}
\end{figure}

\sys{} received an average SUS score of $85.77$, placing the tool in the ``excellent'' usability range, with a system utility score of $83.33$~\cite{brooke1996sus,Bangor_SUS_adjective}.  \autoref{fig:distribution} shows the distribution of responses to the system utility and usability questions. Qualitative feedback further supports this rating. 
\sys{}'s ease of use, efficiency in data narrative composition, and integrated framework were repeatedly commended.
$P11$ described \sys~as ``\textit{something that I could definitely see myself using in my job.}'' 
$P8$ expressed excitement about \sys{}'s potential to address a ``\textit{major hurdle}'' and ``\textit{open the doors to a lot of [non-data-experts] getting a lot closer to data.}'' $P7$ noted that generating text based on the selected data in the charts helped streamline the authoring process: ``\textit{It’s so time-consuming to write insights manually, and the system does it for you, which is a huge time-saver}'' while $P9$ found the overall workflow ``\textit{buttery smooth}''.

Beyond feedback on the overall effectiveness of \sys{}, the evaluation results revealed key insights into the tool's framework and core features, along with areas for improvement:

\noindent\textit{\textbf{The integrated framework is powerful, intuitive, and unique but can benefit from flexible interface layout options.}} 
Participants appreciated the intuitive integrated, bidirectional framework of \sys{}, emphasizing that this functionality allowed for a more seamless transition between data and narrative composition, as $P7$ remarked, ``\textit{The ability to jump from text to visualization and back again without losing context was really powerful.}''  $P9$ added, ``\textit{I loved how the system responded when I selected text--it immediately gave me a chart that matched the narrative, and that felt very smooth.}'' $P8$ described this feature as ``\textit{super powerful}'' and ``\textit{really unique.}'' Participants also noted that the flexibility to begin with either direction significantly enhanced their creative process. For instance, P4 mentioned, ``\textit{I love that I can start with visualization and create text, or start with text and get a chart. It's really versatile and fits the way I think}.'' This flexibility enables users to follow their workflow naturally, whether starting with data or using the narrative to guide analysis. Participants also emphasized \sys{}'s ability to integrate the fragmented authoring process; $P12$ noted, ``\textit{This is so much faster than what my current flow is},'' referring to the frequent switching between different tools. $P2$ concurred, ``\textit{I don't need to switch tools for visualization and text--I can accomplish everything within a single tool.}''
While the integrated framework showed promise, participants expressed diverse preferences for interface layouts that aligned with their mental models. $P7$ expressed a strong preference for our flow-based canvas, stating, ``\textit{I really love this kind of node and link connection way of seeing it... this is so fun.}'' $P2$ remarked that the combination of text and visualization nodes allowed her to ``\textit{iterate and refine the relationship between texts and visualizations.}'' Meanwhile, 
three participants ($P1$, $P9$, $P12$) preferred a \textit{``side-by-side''} or \textit{``left-to-right''} layout to distinctly separate charts and text, preferably with the insight cart ``\textit{in between these two columns}'' [$P9$]. $P5$ advocated for a \textit{``notebook-style''} layout, allowing users to choose their preferred option.

\noindent\textit{\textbf{\textit{Vis-to-text} composition supports efficiency and reduces cognitive workload but demands a more advanced mechanism for filtering and selecting data facts.}}
All participants appreciated the relevance and utility of the generated facts. $P1$ noted, ``\textit{I didn’t come across any instance where the generated facts were inaccurate or out of context. They were spot on, which is crucial for narrative integrity.}'' $P2$ pointed out, ``\textit{the data facts generated are very, very relevant to what I selected}.'' Several participants ($P7$, $P9$, $P13$) highlighted that the generated narratives anchored the storytelling process. $P7$ remarked, \textit{``The facts that were generated really guided me through the process, and I felt like they gave me a clear direction to follow in building my narrative.''} 
However, the sheer volume of information was overwhelming for some participants ($P2$, $P6$, $P12$) --``\textit{There's just so much going on... it’s hard to quickly find what’s important without getting lost in all the details.} [$P6$]'' To address this issue, participants suggested implementing filtering mechanisms that would allow them to prioritize or refine the data facts based on specific needs. $P12$ noted, ``\textit{It would be helpful to have a filter to narrow down the facts based on criteria like time ranges, categories, or trends.}'' Another suggestion was the use of visual glyphs or other non-textual indicators to help visually scan and identify important trends or outliers without needing to read through all the text-based facts. 
Additionally, $P2$ mentioned, ``\textit{It would be great if the data facts were grouped or tiered, so I could drill down from high-level summaries to more detailed information.}'' 
This suggestion underscores the need for a more advanced hierarchical organization of data facts, such as grouping by semantic level~\cite{Lundgard:2022}, complexity, or role. This capability would complement \sys{}'s current management mechanism that categorizes data facts by types and attributes and sorts them within lists based on significance.   

\noindent\textit{\textbf{\textit{Text-to-vis} composition showed promise but also exhibited inconsistent performance due to data limitation.}}
Participants acknowledged the potential of the \textit{text-to-vis} workflow. $P3$ noted that ``\textit{the recommended visualizations are useful in terms of helping me expand my narratives.}''
$P5$ regarded the chart generation feature as ``\textit{the most impressive one.}'' 
$P13$ concurred, describing ``\textit{writing that sentence and having it suggest visualization}'' as their ``\textit{favorite feature}.'' Nevertheless, several participants ($P3$, $P6$, $P11$) encountered challenges in generating meaningful visualizations from text, indicating a limitation in the system's ability to fully capture the user's narrative intent. 
$P6$ explained, ``\textit{It's not always clear what kind of chart will be generated from the text... sometimes I expected a line graph but got something completely different.}'' In response to that disconnect, participants modified the narrative text to explicitly request a specific chart type or define the visual mapping. Moreover, $P11$ felt the recommended visualizations ``\textit{were kind of similar to the charts that already existed.}'' Such inconsistent performance pointed to a current limitation: \sys{} is restricted to performing basic data operations (e.g., filtering) on user-provided or existing datasets. However, the specific charts some participants expected (e.g., a line chart showing a movie director's accumulated gross) often exceeded the scope of the provided datasets or required advanced data manipulation (e.g., data rollup).

\noindent\textit{\textbf{Customization desired for generated visual and textual content.}} 
Besides feedback on the core authoring features, participants highlighted the need for greater customization of the final presentation, encompassing both visual and textual components. For visualizations, they emphasized the importance of flexibility in customizing appearance and formatting, such as color schemes, axis labels, and chart types. $P10$ stated, ``\textit{...customization options would be a big thing for me.}'' Similarly, $P11$ remarked, ``\textit{It would be really useful to have more options to tweak the charts that get generated.}''. Additionally, $P9$ attempted to convert a generated line chart into a bar chart by typing in the text editor, suggesting the potential for a feature to customize charts using natural language commands. 
For textual content, participants ($P2$, $P4$, $P7$, $P12$, $P13$) praised \sys{}'s ability to quickly adjust generated narratives while maintaining their connection to data facts. 
Furthermore, $P8$ suggested incorporating more customizable annotations to better connect visual and textual components, such as overlaying the text on corresponding visuals.

\section{\add{Limitations and Future Implications}}\label{sec:futurework}
The user evaluation reveals key limitations, highlighting areas for improvement and broader applicability.

\noindent\textbf{Further Expansion of the Callout Intent Taxonomy.}
We consider the callout intent taxonomy for visualization a preliminary framework. While the taxonomy covers common chart types and interactions, offering useful guidance for the development of \sys{}, it is not exhaustive. Future work could expand the taxonomy along several dimensions. One apparent avenue is to incorporate a broader repertoire of chart types. We envision two contrasting approaches. For domain-specific visualizations, callout interactions and data facts could be tailored by leveraging conventions and expertise. Alternatively, for dynamically generated visualizations, such as those created with Tableau's ShowMe~\cite{tableau_showme_2024}, a more adaptable and sophisticated approach would be necessary. Another direction could involve additional interaction modalities~\cite{lee_srinivasan_isenberg_stasko_2021}, e.g., sketching~\cite{Lin2023InkSight:Notebooks}. An advantage of alternative interaction modalities is that they do not interfere with interactions used for other analytical tasks. \add{Beyond designing novel callout interactions, future work can explore multiple callout interactions to further disambiguate callout intent, including \textit{compounded callouts} that combine multiple callouts in a single chart, and \textit{chained callouts}, which link callouts across connected charts.}

\noindent\textbf{Lowering Data Barriers for Text-first Authors.}
The study participants underlined a significant limitation: supporting authors with minimal data expertise in creating data-driven articles.
The `text-driven' workflow, while reducing the expertise and effort required to create charts, still requires authors' familiarity with chart interaction and interpretation~\cite{Fu_2023_Morethan}. $P6$ noted that non-data-experts often ``\textit{scout over the facts}'' during authoring and suggested providing these facts along with their \textit{``correlation or causation''} to the narratives. To further simplify this process, we propose extracting data facts pertinent to a focused narrative.
For instance, when an author writes a narrative such as ``\textit{Women have made significant strides in achieving equal representation in the Olympics}'', 
automatically extracting relevant data facts, such as \textit{``In 1900, women made up 2.2\% of Olympic athletes''} and \textit{``The Paris Olympics achieved 50\% female participation''} would assist `text-first' authors in composing data-driven stories. To mitigate potential confirmation bias during this process, we advocate for a more balanced mechanism that incorporates both `supporting' and `refuting' data evidence.

\noindent\textbf{Supporting More Sophisticated Chart Generation.}
We envision two directions to further enhance \sys's capability to generate charts for storytelling. The first involves developing a more advanced data extraction mechanism that can dynamically retrieve and integrate data~\cite{Kezunovic_data_integration} from multiple sources or databases. The second focuses on developing a more robust model capable of proficiently handling complex data manipulations to meet the requirements of the visualization engine. Beyond visualizations that aid data exploration, future research could also explore generating annotated charts~\cite{Lai_2020_automatic_annotation} to accompany textual data facts, providing `visual data facts' to enrich the data story. 

\noindent\textbf{\add{Opportunities and Risks in Expanding Narrative Generation.}}
\add{Another key limitation of \sys{} is its primary focus on assembling together data facts while lacking support for higher-level storylines. This constraint stems from our deliberate prompt design, which restricts text composition to `accurate facts' rather than allowing LLMs to interpret data, thereby leaving interpretation to human authors. However, our user study revealed that while data facts and descriptions serve as foundational building blocks, adding context and high-level takeaways could enrich storytelling by incorporating deeper semantic meaning and interpretation. Future iterations could relax these restrictions, enabling more layered narratives beyond data facts~\cite{Lundgard:2022, bromley2024dashbimodaldataexploration}. For instance, LLMs could be employed to provide context for specific data facts or generate interpretations based on a collection of data points. This approach allows authors to tailor data storytelling for diverse audiences and perspectives, such as low-level data facts for a technical audience and high-level insights for C-level executives. However, relying on LLMs for more creative tasks increases the risk of misinterpretation and bias. Future work should systematically examine the trade-offs between leveraging LLMs for higher-order tasks and mitigating their associated risks.}

\noindent\textbf{\add{Support for Advanced Analysis and Presentation.}}
\add{While \sys{} concentrates on the dual composition of visualization and text, enhancing its framework with advanced data analysis and presentation capabilities is crucial for more insightful and compelling storytelling. We envision this integrated framework as a strong foundation for human-AI collaboration in data storytelling, offering flexible extensibility on both ends. For instance, our flow-based architecture could enable more advanced visual data exploration by connecting multiple visualizations and supporting interactions like brushing and linking, making \sys{} compatible with data analysis frameworks like VisFlow~\cite{YuB17}. With enhanced data processing capabilities, callout interactions could be employed to support \textit{vis-to-vis} composition, i.e., interacting with one chart and recommending others for deeper exploration. Additionally, the auto-established vis-callout-text connections could be adapted to enable more presentation formats (e.g., video~\cite{Shen2023DataInterplay}) and potentially support more expressive chart reconfigurations, such as animated unit visualizations~\cite{Cao_2023_DataParticles}.   
}

\noindent\textbf{\add{Reflection on Integrating LLMs in Data-driven Storytelling.}}
\add{LLMs have the potential to bridge gaps in the multifaceted data storytelling process. Yet, balancing human agency and creativity with automation, efficiency, and accuracy remains both challenging and crucial. We posit that decomposing the workflow into smaller tasks and strategically delegating them is essential to leveraging the complementary strengths effectively. Implementing such a tool requires understanding the pain points and preferences of different user groups (e.g., journalists vs. data analysts), along with a more comprehensive empirical analysis of LLM performance and the risks associated with various task assignments.}

\section{Conclusion}
Creating cohesive, interactive data-driven stories is complex, requiring the integration of visualizations and narrative components. \sys{} addresses these challenges by introducing a framework that supports and integrates both visualization-to-text and text-to-visualization compositions. The tool enables users to ground their narratives in data facts derived from call-out interactions, ensuring stronger alignment between the visualizations and the resulting narrative. By offering both data-first and text-first workflows, the tool provides flexibility for different authoring preferences and supports various presentation formats, such as scrollytelling. A user evaluation suggests that \sys{} streamlines the data storytelling process while offering opportunities for further refinement. The study also highlights a key tension in AI-enabled data narratives: balancing the efficiency of automation with the need for customization. Leveraging AI and computational algorithms to enhance efficiency and reduce workload, while preserving the user's ability to control the narrative structure, tone, and depth of insights could foster more meaningful data-driven stories that faithfully reflect both the data and the author's intent.












\bibliographystyle{eg-alpha-doi} 
\bibliography{bibliography, mendeley_references}

\newcommand{\etalchar}[1]{$^{#1}$}
\begin{thebibliography}{\uppercase{XHRC{\etalchar{*}}18}}

\bibitem[AES05]{Amar_2005_low_level_analytic_activity}
\textsc{Amar R., Eagan J., Stasko J.}:
\newblock {Low-level Components of Analytic Activity in Information
  Visualization}.
\newblock In \emph{IEEE Symposium on Information Visualization, 2005. INFOVIS
  2005.} (2005), pp.~111--117.
\newblock \href {https://doi.org/10.1109/INFVIS.2005.1532136}
  {\path{doi:10.1109/INFVIS.2005.1532136}}.

\bibitem[ARL{\etalchar{*}}16]{amini2016authoring}
\textsc{Amini F., Riche N.~H., Lee B., Monroy-Hernandez A., Irani P.}:
\newblock {Authoring Data-driven Videos with Dataclips}.
\newblock \emph{IEEE Transactions on Visualization and Computer Graphics 23}, 1
  (2016), 501--510.
\newblock \href {https://doi.org/10.1109/TVCG.2016.2598647}
  {\path{doi:10.1109/TVCG.2016.2598647}}.

\bibitem[BKM09]{Bangor_SUS_adjective}
\textsc{Bangor A., Kortum P., Miller J.}:
\newblock {Determining What Individual SUS Scores Mean: Adding an Adjective
  Rating Scale}.
\newblock \emph{J. Usability Studies 4}, 3 (May 2009), 114–123.

\bibitem[BRCP17]{bach2017emerging}
\textsc{Bach B., Riche N.~H., Carpendale S., Pfister H.}:
\newblock {The Emerging Genre of Data Comics}.
\newblock \emph{IEEE Computer Graphics and Applications 37}, 3 (2017), 6--13.
\newblock \href {https://doi.org/10.1109/MCG.2017.33}
  {\path{doi:10.1109/MCG.2017.33}}.

\bibitem[Bro96]{brooke1996sus}
\textsc{Brooke J.}:
\newblock {SUS: A Quick and Dirty Usability Scale}.
\newblock \emph{Usability Evaluation in Industry} (1996).

\bibitem[BS24]{bromley2024dashbimodaldataexploration}
\textsc{Bromley D., Setlur V.}:
\newblock {DASH: A Bimodal Data Exploration Tool for Interactive Text and
  Visualizations}.
\newblock In \emph{2024 IEEE Visualization and Visual Analytics (VIS)} (2024).
\newblock \href {https://doi.org/10.48550/arXiv.2408.01011}
  {\path{doi:10.48550/arXiv.2408.01011}}.

\bibitem[CECX23]{Cao_2023_DataParticles}
\textsc{Cao Y., E J.~L., Chen Z., Xia H.}:
\newblock {DataParticles: Block-Based and Language-Oriented Authoring of
  Animated Unit Visualizations}.
\newblock CHI '23.
\newblock \href {https://doi.org/10.1145/3544548.3581472}
  {\path{doi:10.1145/3544548.3581472}}.

\bibitem[CLA{\etalchar{*}}20]{Chen_2020_StorySynthesis}
\textsc{Chen S., Li J., Andrienko G., Andrienko N., Wang Y., Nguyen P.~H.,
  Turkay C.}:
\newblock {Supporting Story Synthesis: Bridging the Gap between Visual
  Analytics and Storytelling}.
\newblock \emph{IEEE Transactions on Visualization and Computer Graphics 26}, 7
  (7 2020), 2499--2516.
\newblock \href {https://doi.org/10.1109/TVCG.2018.2889054}
  {\path{doi:10.1109/TVCG.2018.2889054}}.

\bibitem[CRS{\etalchar{*}}14]{Carr2014EurographicsEllipsis}
\textsc{Carr H., Rheingans P., Schumann H., Satyanarayan A., Heer J.}:
\newblock {Eurographics Conference on Visualization (EuroVis) 2014 Authoring
  Narrative Visualizations with Ellipsis}.
\newblock \href {https://doi.org/10.1111/cgf.12392}
  {\path{doi:10.1111/cgf.12392}}.

\bibitem[CTL{\etalchar{*}}18]{Chevalier2018}
\textsc{Chevalier F., Tory M., Lee B., van Wijk J., Santucci G., Dörk M.,
  Hullman J.}:
\newblock {From Analysis to Communication: Supporting the Lifecycle of a
  Story}.
\newblock In \emph{Data-Driven Storytelling}, Riche N.~H., Hurter C.,
  Diakopoulos N., Carpendale S., (Eds.). A K Peters/CRC Press, 2018,
  pp.~95--115.

\bibitem[CVTH21]{Conlen2021IdyllArticles}
\textsc{Conlen M., Vo M., Tan A., Heer J.}:
\newblock {Idyll Studio: A Structured Editor for Authoring Interactive {\&}
  Data-Driven Articles}.
\newblock In \emph{ACM Symposium on User Interface Software and Technology} (10
  2021), ACM, pp.~1--12.
\newblock \href {https://doi.org/10.1145/3472749.3474731}
  {\path{doi:10.1145/3472749.3474731}}.

\bibitem[CX22]{Chen_2022_CrossData}
\textsc{Chen Z., Xia H.}:
\newblock {CrossData: Leveraging Text-Data Connections for Authoring Data
  Documents}.
\newblock CHI '22.
\newblock \href {https://doi.org/10.1145/3491102.3517485}
  {\path{doi:10.1145/3491102.3517485}}.

\bibitem[DMN{\etalchar{*}}17]{analyza}
\textsc{Dhamdhere K., McCurley K.~S., Nahmias R., Sundararajan M., Yan Q.}:
\newblock {Analyza: Exploring Data with Conversation}.
\newblock In \emph{International Conference on Intelligent User Interfaces}
  (2017), IUI, pp.~493--504.
\newblock \href {https://doi.org/10.1145/3025171.3025227}
  {\path{doi:10.1145/3025171.3025227}}.

\bibitem[FGB{\etalchar{*}}24]{Fu_UIST24_data_fact_checking}
\textsc{Fu Y., Guo S., Bursztyn V.~S., Hoffswell J., Rossi R., Stasko J.}:
\newblock {``The Data Says Otherwise'' – Towards Automated Fact-checking and
  Communication of Data Claims}.
\newblock In \emph{ACM Symposium on User Interface Software and Technology}
  (2024).
\newblock \href {https://doi.org/10.1145/3654777.3676359}
  {\path{doi:10.1145/3654777.3676359}}.

\bibitem[FS22]{Fu2022SupportingVisualization}
\textsc{Fu Y., Stasko J.}:
\newblock {Supporting Data-Driven Basketball Journalism through Interactive
  Visualization}.
\newblock CHI '22, ACM.
\newblock \href {https://doi.org/10.1145/3491102.3502078}
  {\path{doi:10.1145/3491102.3502078}}.

\bibitem[FS23]{Fu_2023_Morethan}
\textsc{Fu Y., Stasko J.}:
\newblock {More Than Data Stories: Broadening the Role of Visualization in
  Contemporary Journalism}.
\newblock \emph{IEEE Transactions on Visualization and Computer Graphics}
  (2023), 1--20.
\newblock \href {https://doi.org/10.1109/TVCG.2023.3287585}
  {\path{doi:10.1109/TVCG.2023.3287585}}.

\bibitem[GDA{\etalchar{*}}15]{datatone}
\textsc{Gao T., Dontcheva M., Adar E., Liu Z., Karahalios K.~G.}:
\newblock {DataTone: Managing Ambiguity in Natural Language Interfaces for Data
  Visualization}.
\newblock In \emph{ACM Symposium on User Interface Software Technology (UIST)}
  (2015), ACM, pp.~489--500.
\newblock \href {https://doi.org/10.1145/2807442.2807478}
  {\path{doi:10.1145/2807442.2807478}}.

\bibitem[GGC{\etalchar{*}}21]{GadhavePREDICTINGVISUALIZATIONS}
\textsc{Gadhave K., Görtler J., Cutler Z., Nobre C., Deussen O., Meyer M.,
  Phillips J.~M., Lex A.}:
\newblock {Predicting Intent Behind Selections in Scatterplot Visualizations}.
\newblock \emph{Information Visualization 20}, 4 (2021), 207--228.
\newblock \href {https://doi.org/10.1177/14738716211038604}
  {\path{doi:10.1177/14738716211038604}}.

\bibitem[HAW08]{Heer_2008_generalized_selection}
\textsc{Heer J., Agrawala M., Willett W.}:
\newblock {Generalized Selection via Interactive Query Relaxation}.
\newblock In \emph{SIGCHI Conference on Human Factors in Computing Systems}
  (2008), CHI '08, ACM, p.~959–968.
\newblock \href {https://doi.org/10.1145/1357054.1357203}
  {\path{doi:10.1145/1357054.1357203}}.

\bibitem[HD11]{Hullman_2011_narrativevis}
\textsc{Hullman J., Diakopoulos N.}:
\newblock {Visualization Rhetoric: Framing Effects in Narrative Visualization}.
\newblock \emph{IEEE Transactions on Visualization and Computer Graphics 17},
  12 (2011), 2231--2240.
\newblock \href {https://doi.org/10.1109/TVCG.2011.255}
  {\path{doi:10.1109/TVCG.2011.255}}.

\bibitem[{IMD}24]{imdb}
\textsc{{IMDB}}:.
\newblock \url{https://developer.imdb.com/}, 2024.

\bibitem[KP07]{Kezunovic_data_integration}
\textsc{Kezunovic M., Popovic T.}:
\newblock {Substation Data Integration for Automated Data Analysis Systems}.
\newblock In \emph{2007 IEEE Power Engineering Society General Meeting} (2007),
  pp.~1--6.
\newblock \href {https://doi.org/10.1109/PES.2007.386177}
  {\path{doi:10.1109/PES.2007.386177}}.

\bibitem[LKS13]{Lee_2013_SketchStory}
\textsc{Lee B., Kazi R.~H., Smith G.}:
\newblock {SketchStory: Telling More Engaging Stories with Data through
  Freeform Sketching}.
\newblock \emph{IEEE Transactions on Visualization and Computer Graphics 19},
  12 (2013), 2416--2425.
\newblock \href {https://doi.org/10.1109/TVCG.2013.191}
  {\path{doi:10.1109/TVCG.2013.191}}.

\bibitem[LLJ{\etalchar{*}}20]{Lai_2020_automatic_annotation}
\textsc{Lai C., Lin Z., Jiang R., Han Y., Liu C., Yuan X.}:
\newblock {Automatic Annotation Synchronizing with Textual Description for
  Visualization}.
\newblock In \emph{Proceedings of the 2020 CHI Conference on Human Factors in
  Computing Systems} (2020), CHI '20, ACM, p.~1–13.
\newblock \href {https://doi.org/10.1145/3313831.3376443}
  {\path{doi:10.1145/3313831.3376443}}.

\bibitem[LLY{\etalchar{*}}24]{Lin2023InkSight:Notebooks}
\textsc{Lin Y., Li H., Yang L., Wu A., Qu H.}:
\newblock {InkSight: Leveraging Sketch Interaction for Documenting Chart
  Findings in Computational Notebooks}.
\newblock \emph{IEEE Transactions on Visualization and Computer Graphics 30}, 1
  (2024), 944--954.
\newblock \href {https://doi.org/10.1109/TVCG.2023.3327170}
  {\path{doi:10.1109/TVCG.2023.3327170}}.

\bibitem[LRIC15]{Lee_more_than}
\textsc{Lee B., Riche N.~H., Isenberg P., Carpendale S.}:
\newblock {More Than Telling a Story: Transforming Data into Visually Shared
  Stories}.
\newblock \emph{IEEE Comput. Graph. Appl. 35}, 5 (Sep 2015), 84–90.
\newblock \href {https://doi.org/10.1109/MCG.2015.99}
  {\path{doi:10.1109/MCG.2015.99}}.

\bibitem[LS22]{Lundgard:2022}
\textsc{Lundgard A., Satyanarayan A.}:
\newblock {Accessible Visualization via Natural Language Descriptions: A
  Four-Level Model of Semantic Content}.
\newblock \emph{IEEE Transactions on Visualization and Computer Graphics 28}, 1
  (Jan 2022), 1073–1083.
\newblock \href {https://doi.org/10.1109/TVCG.2021.3114770}
  {\path{doi:10.1109/TVCG.2021.3114770}}.

\bibitem[LSIS21]{lee_srinivasan_isenberg_stasko_2021}
\textsc{Lee B., Srinivasan A., Isenberg P., Stasko J.}:
\newblock {Post-WIMP Interaction for Information Visualization}.
\newblock \emph{Foundations and Trends® in Human–Computer Interaction 14}, 1
  (2021), 1–95.
\newblock \href {https://doi.org/10.1561/1100000081}
  {\path{doi:10.1561/1100000081}}.

\bibitem[LSW{\etalchar{*}}21]{LanKineticharts:Design}
\textsc{Lan X., Shi Y., Wu Y., Jiao X., Cao N.}:
\newblock {Kineticharts: Augmenting Affective Expressiveness of Charts in Data
  Stories with Animation Design}.
\newblock \emph{IEEE Transactions on Visualization and Computer Graphics 28}, 1
  (2021), 933--943.
\newblock \href {https://doi.org/10.1109/TVCG.2021.3114775}
  {\path{doi:10.1109/TVCG.2021.3114775}}.

\bibitem[LTW{\etalchar{*}}18]{Liu2018DataAuthoring}
\textsc{Liu Z., Thompson J., Wilson A., Dontcheva M., Delorey J., Grigg S.,
  Kerr B., Stasko J.}:
\newblock {Data Illustrator: Augmenting Vector Design Tools with Lazy Data
  Binding for Expressive Visualization Authoring}.
\newblock In \emph{Conference on Human Factors in Computing Systems -
  Proceedings} (4 2018), vol.~2018-April, ACM.
\newblock \href {https://doi.org/10.1145/3173574.3173697}
  {\path{doi:10.1145/3173574.3173697}}.

\bibitem[LWQ24]{Li2024WhereCollaboration}
\textsc{Li H., Wang Y., Qu H.}:
\newblock {Where Are We So Far? Understanding Data Storytelling Tools from the
  Perspective of Human-AI Collaboration}.
\newblock In \emph{CHI Conference on Human Factors in Computing Systems} (5
  2024), ACM, pp.~1--19.
\newblock \href {https://doi.org/10.1145/3613904.3642726}
  {\path{doi:10.1145/3613904.3642726}}.

\bibitem[LWS{\etalchar{*}}22]{Lan2022NegativeStories}
\textsc{Lan X., Wu Y., Shi Y., Chen Q., Cao N.}:
\newblock {Negative Emotions, Positive Outcomes? Exploring the Communication of
  Negativity in Serious Data Stories}.
\newblock In \emph{Conference on Human Factors in Computing Systems -
  Proceedings} (4 2022), ACM.
\newblock \href {https://doi.org/10.1145/3491102.3517530}
  {\path{doi:10.1145/3491102.3517530}}.

\bibitem[LYZ{\etalchar{*}}23]{Li2023Notable:Notebooks}
\textsc{Li H., Ying L., Zhang H., Wu Y., Qu H., Wang Y.}:
\newblock {Notable: On-the-fly Assistant for Data Storytelling in Computational
  Notebooks}.
\newblock In \emph{CHI Conference on Human Factors in Computing Systems}
  (2023), CHI '23, ACM.
\newblock \href {https://doi.org/10.1145/3544548.3580965}
  {\path{doi:10.1145/3544548.3580965}}.

\bibitem[LZK{\etalchar{*}}22]{latif2021kori}
\textsc{Latif S., Zhou Z., Kim Y., Beck F., Kim N.~W.}:
\newblock {Kori: Interactive Synthesis of Text and Charts in Data Documents}.
\newblock \emph{IEEE Transactions on Visualization and Computer Graphics 28}, 1
  (2022), 184--194.
\newblock \href {https://doi.org/10.1109/TVCG.2021.3114802}
  {\path{doi:10.1109/TVCG.2021.3114802}}.

\bibitem[MMCV23]{Masson2023Charagraph:Paragraphs}
\textsc{Masson D., Malacria S., Casiez G., Vogel D.}:
\newblock {Charagraph: Interactive Generation of Charts for Realtime Annotation
  of Data-Rich Paragraphs}.
\newblock In \emph{Conference on Human Factors in Computing Systems} (4 2023),
  ACM.
\newblock \href {https://doi.org/10.1145/3544548.3581091}
  {\path{doi:10.1145/3544548.3581091}}.

\bibitem[NSS21]{Narechania_2021_NL4DV}
\textsc{Narechania A., Srinivasan A., Stasko J.}:
\newblock {NL4DV: A Toolkit for Generating Analytic Specifications for Data
  Visualization from Natural Language Queries}.
\newblock \emph{IEEE Transactions on Visualization and Computer Graphics 27}, 2
  (2021), 369--379.
\newblock \href {https://doi.org/10.1109/TVCG.2020.3030378}
  {\path{doi:10.1109/TVCG.2020.3030378}}.

\bibitem[NYT]{NYT}
{New York Times}.
\newblock \url{https://nytimes.com}.

\bibitem[PSS23]{Pandey2022MEDLEY:Composition}
\textsc{Pandey A., Srinivasan A., Setlur V.}:
\newblock {MEDLEY: Intent-based Recommendations to Support Dashboard
  Composition}.
\newblock vol.~29, pp.~1135--1145.
\newblock \href {https://doi.org/10.1109/TVCG.2022.3209421}
  {\path{doi:10.1109/TVCG.2022.3209421}}.

\bibitem[RBL{\etalchar{*}}17]{RenChartAccent:Storytelling}
\textsc{Ren D., Brehmer M., Lee B., Höllerer T., Choe E.~K.}:
\newblock {ChartAccent: Annotation for Data-Driven Storytelling}.
\newblock In \emph{2017 IEEE Pacific Visualization Symposium (PacificVis)}
  (2017), pp.~230--239.
\newblock \href {https://doi.org/10.1109/PACIFICVIS.2017.8031599}
  {\path{doi:10.1109/PACIFICVIS.2017.8031599}}.

\bibitem[RHDC18]{riche2018data}
\textsc{Riche N.~H., Hurter C., Diakopoulos N., Carpendale S.} (Eds.):
\newblock \emph{Data-Driven Storytelling}.
\newblock AK Peters Visualization Series. CRC Press, 2018.

\bibitem[RQSR24]{rahman2024exploring}
\textsc{Rahman M.~D., Quadri G.~J., Szafir D.~A., Rosen P.}:
\newblock {Exploring Annotation Taxonomy in Grouped Bar Charts: A Qualitative
  Classroom Study}.
\newblock \emph{Information Visualization} (2024).
\newblock \href {https://doi.org/10.1177/14738716241270247}
  {\path{doi:10.1177/14738716241270247}}.

\bibitem[SBL21]{Sultanum2021LeveragingVizflow}
\textsc{Sultanum N., Bylinskii Z., Liu Z.}:
\newblock {Leveraging Text-Chart Links to Support Authoring of Data-Driven
  Articles with VizFlow}.
\newblock In \emph{Conference on Human Factors in Computing Systems -
  Proceedings} (5 2021), ACM.
\newblock \href {https://doi.org/10.1145/3411764.3445354}
  {\path{doi:10.1145/3411764.3445354}}.

\bibitem[SBT{\etalchar{*}}16]{Setlur2016Eviza:Analysis}
\textsc{Setlur V., Battersby S.~E., Tory M., Gossweiler R., Chang A.~X.}:
\newblock {Eviza: A Natural Language Interface for Visual Analysis}.
\newblock In \emph{Annual Symposium on User Interface Software and Technology}
  (10 2016), ACM, pp.~365--377.
\newblock \href {https://doi.org/10.1145/2984511.2984588}
  {\path{doi:10.1145/2984511.2984588}}.

\bibitem[SCC{\etalchar{*}}23]{Sun2022Erato:Interpolation}
\textsc{Sun M., Cai L., Cui W., Wu Y., Shi Y., Cao N.}:
\newblock {Erato: Cooperative Data Story Editing via Fact Interpolation}.
\newblock \emph{IEEE Transactions on Visualization and Computer Graphics 29}, 1
  (2023), 983--993.
\newblock \href {https://doi.org/10.1109/TVCG.2022.3209428}
  {\path{doi:10.1109/TVCG.2022.3209428}}.

\bibitem[SDES19]{Srinivasan_2019_Voder}
\textsc{Srinivasan A., Drucker S.~M., Endert A., Stasko J.}:
\newblock {Augmenting Visualizations with Interactive Data Facts to Facilitate
  Interpretation and Communication}.
\newblock \emph{IEEE Transactions on Visualization and Computer Graphics 25}, 1
  (2019), 672--681.
\newblock \href {https://doi.org/10.1109/TVCG.2018.2865145}
  {\path{doi:10.1109/TVCG.2018.2865145}}.

\bibitem[SH10]{Segel2010NarrativeData}
\textsc{Segel E., Heer J.}:
\newblock {Narrative Visualization: Telling Stories with Data}.
\newblock \emph{IEEE Transactions on Visualization and Computer Graphics 16}, 6
  (2010), 1139--1148.
\newblock \href {https://doi.org/10.1109/TVCG.2010.179}
  {\path{doi:10.1109/TVCG.2010.179}}.

\bibitem[SH14]{Satyanarayan2014AuthoringEllipsis}
\textsc{Satyanarayan A., Heer J.}:
\newblock {Authoring Narrative Visualizations with Ellipsis}.
\newblock \emph{Computer Graphics Forum 33}, 3 (2014), 361--370.
\newblock \href {https://doi.org/10.1111/cgf.12392}
  {\path{doi:10.1111/cgf.12392}}.

\bibitem[SH24]{Snyder2024DIVI:Visualization}
\textsc{Snyder L.~S., Heer J.}:
\newblock {DIVI: Dynamically Interactive Visualization}.
\newblock \emph{IEEE Transactions on Visualization and Computer Graphics 30}, 1
  (1 2024), 403--413.
\newblock \href {https://doi.org/10.1109/TVCG.2023.3327172}
  {\path{doi:10.1109/TVCG.2023.3327172}}.

\bibitem[SKH{\etalchar{*}}23]{Shin_2023_Roslingifier}
\textsc{Shin M., Kim J., Han Y., Xie L., Whitelaw M., Kwon B.~C., Ko S.,
  Elmqvist N.}:
\newblock {Roslingifier: Semi-Automated Storytelling for Animated
  Scatterplots}.
\newblock \emph{IEEE Transactions on Visualization and Computer Graphics 29}, 6
  (2023), 2980--2995.
\newblock \href {https://doi.org/10.1109/TVCG.2022.3146329}
  {\path{doi:10.1109/TVCG.2022.3146329}}.

\bibitem[Sof24]{tableau_showme_2024}
\textsc{Software T.}:
\newblock {Show Me: Automatic Presentation for Visual Analysis}.
\newblock Online Resource, 2024.
\newblock Tableau Software.
\newblock URL: \url{https://www.tableau.com/showme}.

\bibitem[SS18]{orko}
\textsc{Srinivasan A., Stasko J.}:
\newblock {Orko: Facilitating Multimodal Interaction for Visual Exploration and
  Analysis of Networks}.
\newblock \emph{IEEE Transactions on Visualization and Computer Graphics 24}, 1
  (2018), 511--521.
\newblock \href {https://doi.org/10.1109/TVCG.2017.2745219}
  {\path{doi:10.1109/TVCG.2017.2745219}}.

\bibitem[SS23a]{Srinivasan2023BOLT:Authoring}
\textsc{Srinivasan A., Setlur V.}:
\newblock {BOLT: A Natural Language Interface for Dashboard Authoring}.
\newblock In \emph{Eurographics Conference on Visualization} (2023).
\newblock URL: \url{https://api.semanticscholar.org/CorpusID:267751137}.

\bibitem[SS23b]{Sultanum2023DataTales:Articles}
\textsc{Sultanum N., Srinivasan A.}:
\newblock {DataTales: Investigating the use of Large Language Models for
  Authoring Data-Driven Articles}.
\newblock In \emph{2023 IEEE Visualization and Visual Analytics (VIS)} (2023),
  pp.~231--235.
\newblock \href {https://doi.org/10.1109/VIS54172.2023.00055}
  {\path{doi:10.1109/VIS54172.2023.00055}}.

\bibitem[SXS{\etalchar{*}}21]{Shi_2021_Colliope}
\textsc{Shi D., Xu X., Sun F., Shi Y., Cao N.}:
\newblock {Calliope: Automatic Visual Data Story Generation from a
  Spreadsheet}.
\newblock \emph{IEEE Transactions on Visualization and Computer Graphics 27}, 2
  (2021), 453--463.
\newblock \href {https://doi.org/10.1109/TVCG.2020.3030403}
  {\path{doi:10.1109/TVCG.2020.3030403}}.

\bibitem[SZZW24]{Shen2023DataInterplay}
\textsc{Shen L., Zhang Y., Zhang H., Wang Y.}:
\newblock {Data Player: Automatic Generation of Data Videos with
  Narration-Animation Interplay}.
\newblock \emph{IEEE Transactions on Visualization \& Computer Graphics 30}, 01
  (jan 2024), 109--119.
\newblock \href {https://doi.org/10.1109/TVCG.2023.3327197}
  {\path{doi:10.1109/TVCG.2023.3327197}}.

\bibitem[tab]{tableaupublic}
{Tableau Public}.
\newblock \url{https://public.tableau.com/}.

\bibitem[TLS21]{Thompson2021DataGraphics}
\textsc{Thompson J.~R., Liu Z., Stasko J.}:
\newblock {Data Animator: Authoring Expressive Animated Data Graphics}.
\newblock In \emph{CHI Conference on Human Factors in Computing Systems}
  (2021), CHI '21, ACM.
\newblock \href {https://doi.org/10.1145/3411764.3445747}
  {\path{doi:10.1145/3411764.3445747}}.

\bibitem[WSZ{\etalchar{*}}20]{Wang_2020DataShot}
\textsc{Wang Y., Sun Z., Zhang H., Cui W., Xu K., Ma X., Zhang D.}:
\newblock {DataShot: Automatic Generation of Fact Sheets from Tabular Data}.
\newblock \emph{IEEE Transactions on Visualization and Computer Graphics 26}, 1
  (2020), 895--905.
\newblock \href {https://doi.org/10.1109/TVCG.2019.2934398}
  {\path{doi:10.1109/TVCG.2019.2934398}}.

\bibitem[XHRC{\etalchar{*}}18]{Xia2018Dataink:Drawing}
\textsc{Xia H., Henry~Riche N., Chevalier F., De~Araujo B., Wigdor D.}:
\newblock {{DataInk}: Direct and Creative Data-Oriented Drawing}.
\newblock In \emph{CHI Conference on Human Factors in Computing Systems}
  (2018), CHI '18, ACM, p.~1–13.
\newblock \href {https://doi.org/10.1145/3173574.3173797}
  {\path{doi:10.1145/3173574.3173797}}.

\bibitem[YKSJ07]{Yi2007TowardVisualization}
\textsc{Yi J.~S., Kang Y.~a., Stasko J., Jacko J.}:
\newblock {Toward a Deeper Understanding of the Role of Interaction in
  Information Visualization}.
\newblock \emph{IEEE Transactions on Visualization and Computer Graphics 13}, 6
  (2007), 1224--1231.
\newblock \href {https://doi.org/10.1109/TVCG.2007.70515}
  {\path{doi:10.1109/TVCG.2007.70515}}.

\bibitem[YS17]{YuB17}
\textsc{{Yu} B., {Silva} C.~T.}:
\newblock {VisFlow} - {Web-based Visualization Framework for Tabular Data with
  a Subset Flow Model}.
\newblock \emph{IEEE Transactions on Visualization and Computer Graphics 23}, 1
  (2017), 251--260.
\newblock \href {https://doi.org/10.1109/TVCG.2016.2598497}
  {\path{doi:10.1109/TVCG.2016.2598497}}.

\bibitem[YS20]{Yu2020FlowSense:System}
\textsc{Yu B., Silva C.~T.}:
\newblock {FlowSense: A Natural Language Interface for Visual Data Exploration
  within a Dataflow System}.
\newblock \emph{IEEE Transactions on Visualization and Computer Graphics 26}, 1
  (1 2020), 1--11.
\newblock \href {https://doi.org/10.1109/TVCG.2019.2934668}
  {\path{doi:10.1109/TVCG.2019.2934668}}.

\bibitem[YXL{\etalchar{*}}22]{Yang_2022_data_story_structure}
\textsc{Yang L., Xu X., Lan X., Liu Z., Guo S., Shi Y., Qu H., Cao N.}:
\newblock {A Design Space for Applying the Freytag's Pyramid Structure to Data
  Stories}.
\newblock \emph{IEEE Transactions on Visualization and Computer Graphics 28}, 1
  (2022), 922--932.
\newblock \href {https://doi.org/10.1109/TVCG.2021.3114774}
  {\path{doi:10.1109/TVCG.2021.3114774}}.

\bibitem[ZE23]{zhao2023stories}
\textsc{Zhao Z., Elmqvist N.}:
\newblock {The Stories We Tell About Data: Surveying Data-driven Storytelling
  Using Visualization}.
\newblock \emph{IEEE Computer Graphics and Applications 43}, 4 (2023), 97--110.
\newblock \href {https://doi.org/10.1109/MCG.2023.3269850}
  {\path{doi:10.1109/MCG.2023.3269850}}.

\bibitem[ZHC24]{Zhou2024Epigraphics:Authoring}
\textsc{Zhou T., Huang J., Chan G. Y.-Y.}:
\newblock {Epigraphics: Message-Driven Infographics Authoring}.
\newblock ACM (ACM), pp.~1--18.
\newblock \href {https://doi.org/10.1145/3613904.3642172}
  {\path{doi:10.1145/3613904.3642172}}.

\end{thebibliography}





\end{document}